\documentclass[aps,10pt,prd,a4paper,twocolumn,notitlepage,nofootinbib,showpacs]{revtex4-1}
\makeatother
\usepackage{amstext}
\usepackage[latin9]{inputenc}
\usepackage[T1]{fontenc}
\usepackage{amsmath}
\usepackage{graphicx}
\usepackage{color}
\usepackage[normalem]{ulem} 
\usepackage{caption}
\begin{document}
 
\title{On extensions of the Penrose inequality  with angular momentum  }

\author{Wojciech Kulczycki}
\author{Edward Malec}
\affiliation{Instytut Fizyki im.~Mariana Smoluchowskiego, Uniwersytet Jagiello\'nski, {\L}ojasiewicza 11, 30-348 Krak\'{o}w, Poland} 

\begin{abstract}
	We  numerically investigate the validity of recent modifications of the Penrose inequality that include angular momentum.   Formulations   expressed in terms of asymptotic mass and asymptotic angular momentum are contradicted. We analyzed   numerical solutions describing polytropic stationary toroids around spinning black holes. 	 
\end{abstract}

\pacs{04.20.-q, 04.25.Nx, 04.40.Nr, 95.30.Sf}
\maketitle 

\section{Introduction}
The Cosmic Censorship Hypothesis, originally formulated  by Roger Penrose more than half century ago \cite{Penrose1969}, can be understood as a statement that classical general relativity is self-contained, when describing regions exterior to black holes.   Penrose  has  
argued  that the Cosmic Censorship Hypothesis cannot be true       if a   body collapsing to a black hole fails to  satisfy
  the inequality
\begin{equation}M_\mathrm{ADM} \ge \sqrt{\frac{S}{16\pi }}, \label{1}
\end{equation}
where $S$ is the area of the outermost apparent horizon that surrounds the body and $M_\mathrm{ADM}$ is the asymptotic mass of the spacetime  \cite{Penrose1973}. (Herein and in what follows we always assume asymptotic flatness of a spacetime.)  Failure in satisfying of (\ref{1}) would imply the existence of a ``naked singularity'' and a loss of predictability in a collapsing system.
There is probably no exaggeration in saying that this idea has shaped the development of classical and mathematical general relativity in the last five decades.

The Penrose inequality has been proven or checked numerically in a number of special cases --- conformally flat systems with matter \cite{KMS}, Brill gravitational waves in  Weyl geometries \cite{Koc} and    various foliations of spherically symmetric systems \cite{MNOM, Hayward}. Most remarkably, it was proven in the  important     so-called Riemannian case, when the apparent horizon coincides with a minimal surface \cite{Huisken, Huisken1, Bray}.  There exist  scenarios for a general proof \cite{Frauendiener,MMS}, but there are no easy prospects for their implementation.  For more information see specialized reviews, for instance \cite{KM, Mars}.

Christodoulou introduced the concept of an irreducible mass $M_\mathrm{irr}=\sqrt{S/(16\pi )}$ \cite{Christodoulu}, where $S$ is the area of the event horizon. It appears that for the Kerr spacetime endowed with the asymptotic mass $M_\mathrm{ADM}$ and the angular momentum  $J_\mathrm{ADM}$ one has the relation $M^2_\mathrm{ADM}=M^2_\mathrm{irr}+\frac{  J^2_\mathrm{ADM}}{4M^2_\mathrm{irr}} =\frac{S}{16\pi }+\frac{4\pi J^2_\mathrm{ADM}}{S}$.  An analog of this formula might be used  in order to define the quasilocal mass of a black hole (assuming that the linear momentum of the black hole vanishes) in terms of its area and quasilocal angular momentum $J_\mathrm{BH}$:
$M_\mathrm{BH} =\sqrt{M^2_\mathrm{irr}+\frac{  J_\mathrm{BH}^2}{4M^2_\mathrm{irr}}} =\sqrt{\frac{S}{16\pi }+\frac{4\pi J_\mathrm{BH}^2}{S}}$.
This concept of the quasilocal mass of a black hole    is commonly used in the literature.

There exist formulations of the    Penrose inequality that involve the asymptotic mass  and quasilocal angular momentum \cite{Mars,Dain_Gabach},  
\begin{equation}M_\mathrm{ADM} \ge \left(\frac{S}{16\pi }+\frac{4\pi J_\mathrm{BH}^2}{S}\right)^{1/2}; \label{2}
\end{equation}
here $S$ and $J_\mathrm{BH}$  are the area  and quasilocal  angular momentum of the outermost apparent horizon.  Anglada \cite{Anglada} and Khuri \cite{Khuri} have proved other  versions of  (\ref{2}) under  the assumption of axial symmetry, 

\begin{equation}
M_\mathrm{ADM} \ge \left( \frac{S}{16\pi }+\frac{  J_\mathrm{BH}^2}{\tilde R^2(S)}\right)^{1/2};
 \label{3}
\end{equation}
where   $\tilde R(S) $  is some linear measure of the boundary of the black hole.

In what follows we shall study numerically the validity of (\ref{2}) and related inequalities in a class of stationary configurations consisting of a black hole and a torus. The order 
is following. Next Section contains a short description of equations and relevant quantities. Section 3 gives a concise summary of the numerical procedure. The main results are  reported in Section 4.   We conclude the paper with a summary.

\section{Equations}

We assume a \textit{stationary} metric of the form 
\begin{align}
ds^{2} & =-\alpha^{2}{dt}^{2}+r^{2}\sin^{2}\theta\psi^{4}\left(d\varphi+\beta dt\right)^{2}\nonumber \\
 & \qquad+\psi^{4}e^{2q}\left(dr^{2}+r^{2}d\theta^{2}\right). \label{metric}
\end{align}
Here $t$ is the time coordinate, and $r$, $\theta$, $\varphi$
are spherical coordinates. In  this paper the gravitational constant $G=1$ and the
speed of light $c=1$. We assume axial symmetry and employ the stress-momentum
tensor 
\[
T^{\alpha\beta}=\rho hu^{\alpha}u^{\beta}+pg^{\alpha\beta},
\]
where $\rho$ is the baryonic rest-mass density, $h$ is the specific
enthalpy, and $p$ is the pressure. Metric functions $\alpha $,
$\psi $, $q $ and $\beta $ in \eqref{metric}
depend on $r$ and $\theta$ only.

 The forthcoming   Einstein equations  have been found in \cite{MSH} and checked by authors of \cite{prd2018a, prd2018b}; the present formulation follows closely the description of \cite{prd2018a, prd2018b}.

Below $K_{ij}$ denotes the extrinsic curvature of the $t = \mathrm{const}$ hypersurface. The conformal extrinsic curvature $\hat K_{ij}$ is defined as $\hat K_{ij} = \psi^2 K_{ij}$. The only nonzero component $\beta$ of the shift vector  is separated into two parts, $\beta = \beta_\mathrm{K} + \beta_\mathrm{T}$.  Here   $\beta_\mathrm{T}$ is a part of the shift vector related to the rotating torus \cite{MSH}. Functions $\beta_\mathrm{K}$ and $\beta_\mathrm{T}$ are determined as follows. The nonzero components of $\hat K_{ij}$ can be written in the form
\[ \hat K_{r \varphi} = \frac{H_\mathrm{E} \sin^2 \theta}{r^2} + \frac{\psi^6}{2 \alpha} r^2 \sin^2 \theta \partial_r \beta_\mathrm{T}, \]
\[ \hat K_{\theta \varphi} = \frac{H_\mathrm{F} \sin \theta}{r} + \frac{\psi^6}{2 \alpha} r^2 \sin^2 \theta \partial_\theta \beta_\mathrm{T}. \] 
As in \cite{MSH}, we choose the functions $H_\mathrm{E}$ and $H_\mathrm{F}$ to be expressed by the formulae obtained for the Kerr metric of mass $m$ and the spin parameter $a$, written in the form (\ref{metric}). In explicit terms they read \cite{BrandtSeidel1995}
\[ H_\mathrm{E} = \frac{m a \left[ (r_\mathrm{K}^2 - a^2) \Sigma_\mathrm{K} + 2 r_\mathrm{K}^2 (r_\mathrm{K}^2 + a^2) \right]}{\Sigma_\mathrm{K}^2}, \]
\[ H_\mathrm{F} = - \frac{2 m a^3 r_\mathrm{K} \sqrt{r_\mathrm{K}^2 - 2 m r_\mathrm{K} + a^2} \cos \theta \sin^2 \theta}{\Sigma_\mathrm{K}^2}, \]
where
\[ r_\mathrm{K} = r \left( 1 + \frac{m}{r} + \frac{m^2 - a^2}{4 r^2} \right), \]
and
\[ \Sigma_\mathrm{K} = r_\mathrm{K}^2 + a^2 \cos^2 \theta. \]
It appears  that for the Kerr metric one has
\[ \hat K_{r \varphi} = \frac{H_\mathrm{E} \sin^2 \theta}{r^2} \]
and
\[ \hat K_{\theta \varphi} = \frac{H_\mathrm{F} \sin \theta}{r}. \]
The function $\beta_\mathrm{K}$ has to be computed from the relation \cite{MSH}
\begin{equation}
\frac{\partial \beta_\mathrm{K}}{\partial r} = \frac{2 H_\mathrm{E} \alpha}{r^4 \psi^6}. 
\label{eqbetak}
\end{equation}
The function $\beta_\mathrm{T}$, with suitable boundary conditions (see next Section) is found from the differential equation (\ref{46}).

In what follows we apply the puncture method as implemented in \cite{MSH}. Let $\Phi = \alpha \psi$ and  assume the puncture at $r = 0$. Define $r_\mathrm{s} = \frac{1}{2}\sqrt{m^2 - a^2}$, and
\[ \psi = \left(1 + \frac{r_\mathrm{s}}{r} \right) e^\phi, \quad \Phi = \left(1 - \frac{r_\mathrm{s}}{r} \right)e^{-\phi} B. \]
Then the  surface $r = r_\mathrm{s}$ is an apparent horizon.

Einstein equations read
\begin{widetext}
\begin{subequations}
\label{main_sys}
\begin{eqnarray}
\left[ \partial_{rr} + \frac{1}{r } \partial_r  + \frac{1}{r^2} \partial_{\theta \theta}  \right] q & = & S_q, \label{47}\\
\left[ \partial_{rr} + \frac{2 r  }{r^2 - r_\mathrm{s}^2} \partial_r + \frac{1}{r^2} \partial_{\theta \theta} + \frac{  \cot{\theta}}{r^2}  \partial_\theta \right] \phi & = & S_\phi, \label{44} \\
\left[ \partial_{rr} + \frac{3 r^2 +  r_\mathrm{s}^2}{r(r^2 - r_\mathrm{s}^2)} \partial_r + \frac{1}{r^2} \partial_{\theta \theta} + \frac{2 \cot{\theta}}{r^2}  \partial_\theta \right] B & = & S_B, \label{45} \\
\left[ \partial_{rr} + \frac{4 r^2 - 8 r_\mathrm{s} r + 2 r_\mathrm{s}^2}{r(r^2 - r_\mathrm{s}^2)} \partial_r + \frac{1}{r^2} \partial_{\theta \theta} + \frac{3 \cot{\theta}}{r^2}  \partial_\theta \right]  \beta_\mathrm{T} & = & S_{\beta_\mathrm{T}},  \label{46}
\end{eqnarray}
 \end{subequations}
where source terms $S_\phi, S_B, S_{\beta_\mathrm{T}}, S_q $ are: 
\begin{subequations}
\begin{flalign}
&S_q  =  -8 \pi e^{2q} \left( \psi^4 p - \frac{\rho h u_\varphi^2}{r^2 \sin^2 \theta} \right) + \frac{3 A^2}{\psi^8} + 2 \left[ \frac{r - r_\mathrm{s}}{r(r + r_\mathrm{s})} \partial_r + \frac{\cot \theta}{r^2} \partial_\theta \right] b + \left[ \frac{8 r_\mathrm{s}}{r^2 - r_\mathrm{s}^2} + 4 \partial_r (b - \phi) \right] \partial_r \phi  \nonumber\\
&+ \frac{4}{r^2} \partial_\theta \phi \partial_\theta (b - \phi),   \label{5001}\\
&
S_\phi = 
- 2 \pi e^{2q} \psi^4 \left[ \rho_\mathrm{H} - p + \frac{\rho h u_\varphi^2}{\psi^4 r^2 \sin^2 \theta} \right] - \frac{A^2}{\psi^8}  - \partial_r\phi \partial_r b - \frac{1}{r^2} \partial_\theta \phi \partial_\theta b - \frac{1}{2} \left[ \frac{r - r_\mathrm{s}}{r (r + r_\mathrm{s})} \partial_r b + \frac{\cot \theta}{r^2} \partial_\theta b \right],\label{5002} \\
&
S_B = 16 \pi B e^{2q} \psi^4 p, \label{5003}\\
&
S_{\beta_\mathrm{T}} = \frac{16 \pi \alpha e^{2q} \tilde J}{r^2 \sin^2 \theta} - 8 \partial_r \phi \partial_r \beta_\mathrm{T} + \partial_r b \partial_r \beta_\mathrm{T} - 8 \frac{\partial_\theta \phi \partial_\theta \beta_\mathrm{T}}{r^2} +\frac{\partial_\theta b \partial_\theta \beta_\mathrm{T}}{r^2}\label{5004}
\end{flalign}
\end{subequations}
 
\end{widetext}
and
\[ A^2 = \frac{\hat K^2_{r \varphi}}{r^2 \sin^2 \theta} + \frac{\hat K^2_{\theta \varphi}}{r^4 \sin^2 \theta}, \]
\[ \rho_\mathrm{H} = \rho h (\alpha u^t)^2 - p, \]
\[ \tilde J = \rho h \alpha u^t u_\varphi, \]
\[ B = e^b. \]

In the rest of this paper we will deal with polytropes $p(\rho)=K\rho^{\gamma }$.
Then one has the specific enthalpy 
\[
h(\rho)=1+ \frac{\gamma p}{(\gamma -1) \rho}.
\]
The 4-velocity $(u^{\alpha})=(u^{t},0,0,u^{\varphi})$ is normalized,
$g_{\alpha\beta}u^{\alpha}u^{\beta}=-1$. The coordinate angular velocity
reads 
\begin{equation}
\Omega=\frac{u^{\varphi}}{u^{t}}.\label{Omega_def}
\end{equation}
  We  define the angular momentum per unit
inertial mass $\rho h$ \cite{FM} 
\begin{equation}
j\equiv u_{\varphi}u^{t}.\label{j_def}
\end{equation}
It is  known since early 1970's that general-relativistic Euler equations are solvable under the condition that $j\equiv j(\Omega)$ \cite{Butterworth_Ipser,Bardeen_1970}.
Within the fluid region, the Euler equations $\nabla_{\mu}T^{\mu\nu}=0$
can be integrated,
\begin{equation}
\int j(\Omega) d\Omega+\ln\left(\frac{h}{u^{t}}\right)=C,\label{uf}
\end{equation}
where $C$ is a constant.

We shall use in this paper  the rotation laws obtained in \cite{KM2020}
\begin{eqnarray}
\label{momentum}
j(\Omega ) &\equiv &-\frac{1}{1-3\delta}\frac{d}{d\Omega} \ln \left(  1-  ( a\Omega )^2\right. \nonumber\\
&& \left.-   \kappa    w^{1- \delta }\Omega^{1+\delta }(1- a\Omega )^{1-\delta} \right) .
\end{eqnarray}
Here $J_\mathrm{BH}$ and  $a=J_\mathrm{BH}/m$ are the angular momentum and the spin parameter of the central black hole, respectively.    $\delta $ is a free  parameter and $\kappa =\frac{1-3\delta }{1+\delta }$. Let us remark, that (\ref{momentum}) supplements former rotation recipes that have been formulated in  \cite{MM, APP2015} and (in the case of the Keplerian rotation around spinning black holes) \cite{prd2018a, prd2018b}.
   
The Keplerian rotation corresponds to the parameter 
$\delta =-1/3$ and $\kappa =3$.

The rotation curves --- angular velocities as functions of spatial coordinates $\Omega(r,\theta )$ --- can  be recovered from   Eq.\ (\ref{j_def}),
\begin{equation}
j(\Omega)=\frac{V^{2}}{\left(\Omega+\beta\right)\left(1-V^{2}\right)}.\label{rotation_law}
\end{equation}
Here  the squared  linear velocity is given by
\[
V^{2}=r^{2}\sin^{2}\theta\left(\Omega+\beta\right)^{2}\frac{\psi^{4}}{\alpha^{2}}.
\]

 The central black hole is defined by    the puncture method \cite{BrandtSeidel1995}. The   black hole is surrounded by a minimal two-surface $S $ (the horizon)  embedded in a fixed hypersurface of constant time, and
located at  $r=r_{\mathrm{s}} = \sqrt{m^2 -   a^2}/2$, where $m$ is a mass parameter. Its area is denoted as $S$ and its angular momentum $J_{\mathrm{BH}}$ follows from the Komar expression
\begin{equation}
J_{\mathrm{BH}}=\frac{1}{4}\int_{0}^{\pi/2}\frac{r^{4}\psi^{6}}{\alpha}\partial_{r}\beta\sin^{3}\theta d\theta.
\end{equation}
We would like to point that  the angular momentum is given rigidly on the event horizon $S $, in terms of 
   data taken from the Kerr solution  and
independently of the content of mass in a torus,  $J_\mathrm{BH} = m   a$ \cite{MSH}. The mass of the black
hole is then defined in terms of its area and the angular momentum,
\begin{equation}
M_{\mathrm{BH}}=\sqrt{\frac{S}{16\pi }+\frac{4\pi J_{\mathrm{BH}}^{2}}{S}}.
\label{PI}
\end{equation}

 Asymptotic (total) mass $M_\mathrm{ADM}$  and angular momentum $J_\mathrm{ADM}$ can be defined as apropriate Arnowitt-Deser-Misner charges, and they can be computed by means of corresponding volume integrals \cite{MSH}.
 Thus we have 
 \begin{eqnarray}
\label{j1}
M_\mathrm{ADM}&=&\sqrt{m^2-a^2}-2\int_{r_\mathrm{s}}^\infty \left(r^2-r^2_\mathrm{s}\right)dr\int_0^{\pi/2}\sin \theta d\theta S_\phi, \nonumber\\
J_\mathrm{T} &=& 4 \pi \int_{r_\mathrm{s}}^\infty r^2 dr \int_0^{\pi/2} \sin \theta d \theta \rho \alpha u^t \psi^6 e^{2 q} h u_\varphi, \nonumber\\
J_\mathrm{ADM}&=&J_\mathrm{BH}+J_\mathrm{T}.
\end{eqnarray}
Here $J_\mathrm{T}$ is the angular momentum deposited within the torus.
A circumferential radius corresponding to the coordinate  $r=\mathrm{const}$ on the symmetry plane $\theta =\pi /2$
is given by
\begin{equation}
r_\mathrm{C}(r)=r\psi^2(r,\theta =\pi/2).
\label{rad}
\end{equation}
  One can define an alternative mass of the apparent horizon in terms of  $r_\mathrm{C}(r)$,
  \begin{equation}
  M_\mathrm{C}\equiv   \frac{r_\mathrm{C}(r_\mathrm{s})} {2}.
  \label{mc}
  \end{equation}
In  the Kerr spacetime one has exactly $M_\mathrm{C}=M_\mathrm{BH}=m$. It is known that in numerically obtained spacetimes with gaseous toroids the first equality holds with a   good accuracy, albeit depending on spin \cite{MSH, prd2018a,prd2018b}. The second equality is true only approximately for relatively light disks, and it is not true for heavy tori.

\section{Description of numerics}

The numerical method bases on  \cite{MSH} and it has been presented in more details in \cite{prd2018b}. Here we use a more general rotation law   (\ref{momentum}) and   different linear algebra routines ---  the PARDISO library \cite{pardiso} instead of LAPACK \cite{lapack}. In what follows we briefly summarize the main points.

The solutions are found iteratively. In each iteration  one solves the Einstein equations (\ref{main_sys}) and (\ref{5001})-(\ref{5004}), Eq. (\ref{eqbetak}) and   two hydrodynamic equations: (\ref{uf}) and (\ref{rotation_law}). Eqs. (\ref{main_sys}) (with their source terms (\ref{5001})-(\ref{5004})) are solved using a fixed-point method (we use the PARDISO library \cite{pardiso}) with respect to functions $\phi$, $B$, $\beta_{\mathrm{T}}$ and $q$. The function $\beta_\mathrm{K}$ is computed by integration of Eq.\ (\ref{eqbetak}). Eqs. (\ref{uf}) and (\ref{rotation_law}) are used to calculate the specific enthalpy $h$ and the angular velocity $\Omega$, respectively. Constants ($C$ and $w$) that appear in these two equations are computed by solving them at boundary points $(r_1,\pi/2)$ and $(r_2,\pi/2)$, using the Newton-Raphson method.  Here $r_1$ and $r_2$ are the values of the inner and outer radii of the torus, respectively; they are given a priori.  

The free hydrodynamic data consist of the maximal baryonic density $\rho_\mathrm{max}$ and the polytropic index $\gamma $. We assume from now on $\gamma =4/3$. The baryonic density $\rho$ is calculated from the specific enthalpy $h$ using the polytropic formula
\[
\rho=\left[\frac{h-1}{4K } \right]^{\frac{1}{1/3}}.
\]
This yields (in each iteration) the constant $K$ as a function of the maximal value of the specific enthalpy $h_\mathrm{max}$ and  $\rho_\mathrm{max}$
\[
K= \frac{h_\mathrm{max}-1}{4\rho_\mathrm{max}^{1/3}}.
\]
We have assumed   axial and equatorial symmetry and the puncture method with the puncture at $r=r_\mathrm{s}\equiv\sqrt{m^2-a^2}/2$ . We should add that the mass $m$ and the spin $a$   (which appears also in (\ref{momentum}))
are given apriori. Thus it suffices that the numerical grid   covers the region $r_\mathrm{s}\le r<  \infty$ and $0\le\theta\le\pi/2$ with suitable boundary conditions \cite{MSH}. We have at the equator  ($\theta=\pi/2$): $\partial_{\theta}\phi=\partial_{\theta}B=\partial_{\theta}\beta_{\mathrm{T}}=\partial_{\theta}q=0$. The regularity conditions along the axis ($\theta=0$) read: $\partial_{\theta}\phi=\partial_{\theta}B=\partial_{\theta}\beta_{\mathrm{T}}=0$.  It is required that $q(\theta=0)=0$; this is due to the local flatness of the metric. The puncture formalism implies the boundary conditions at the horizon $r=r_\mathrm{s}$: $\partial_{r}\phi=\partial_{r}B=\partial_{r}\beta_{\mathrm{T}}=\partial_{r}q=0$ and $ \partial_{rr}\beta_{\mathrm{T}}=\partial_{rrr}\beta_{\mathrm{T}}=0$. These last two conditions on $\beta_\mathrm{T}$ follow from a careful analysis of  Eq. (\ref{46})  that yields     stringent conditions at $r=r_{\mathrm{s}}$  \cite{MSH}.  

At the outer boundary of the numerical  domain   the boundary conditions are obtained from the multipole expansion and the conditions of asymptotic flatness. Thus we have for $r\to\infty$:
\begin{eqnarray}
\phi & \sim & \frac{M_{1}}{2r}, \ \ \ \ \ B  \sim 1-\frac{B_{1}}{r^{2}}, \nonumber \\
\beta_{\mathrm{T}} & \sim & -\frac{2J_{1}}{r^{3}},  \ \ \ \ \ q  \sim  \frac{q_{1}\sin^{2}\theta}{r^{2}}.\label{boundinf}
\end{eqnarray}
Herein the constants $M_{1}$, $B_{1}$, $J_{1}$ and $q_{1}$ are given by
\begin{equation}
M_{1}=-2\int_{r_{\mathrm{s}}}^{\infty}(r^{2}-r_{\mathrm{s}}^{2})dr\int_{0}^{\pi/2}\sin\theta d\theta S_{\phi},\label{m1}
\end{equation}
\begin{equation}
B_{1}=\frac{2}{\pi}\int_{r_{\mathrm{s}}}^{\infty}dr\frac{(r^{2}-r_{\mathrm{s}}^{2})^{2}}{r}\int_{0}^{\pi/2}d\theta\sin^{2}\theta S_{B},\label{b1}
\end{equation}
\begin{equation}
J_{1}=4\pi\int_{r_{\mathrm{s}}}^{\infty}r^{2}dr\int_{0}^{\pi/2}\sin\theta d\theta\rho\alpha u^{t}\psi^{6}e^{2q}hu_{\varphi},\label{j1}
\end{equation}
\begin{eqnarray}
q_{1} & = & \frac{2}{\pi}\int_{r_{\mathrm{s}}}^{\infty}drr^{3}\int_{0}^{\pi/2}d\theta\cos(2\theta)S_{q}\nonumber \\
 &  & -\frac{4}{\pi}r_{\mathrm{s}}^{2}\int_{0}^{\pi/2}d\theta\cos(2\theta)q(r_{\mathrm{s}},\theta).\label{q1}
\end{eqnarray}
Finally, we add that the Kerr solution  emerges in our method as a vacuum limit  $\rho_\mathrm{max}\rightarrow 0$.

\section{The Penrose inequality in stationary black hole-torus systems}

In the rest of the paper we always assume that $\Omega>0$ and the mass parameter $m=1$. Corotating disks have $a>0$, while counterrotating disks have negative spins: $a<0$.  
The    disk's boundaries are numerically defined by the condition that the specific enthalpy is
$h=1 $. 
The results of numerical calculations are provided in the forthcoming Table.  We shall describe its content in   more detail in the second part of this Section, when referring to new 
proposals of Penrose-type inequalities. In the first part  we will refer to canonical versions.  

\subsection{On   inequalities (\ref{1}) and (\ref{2})}
The mass $M_\mathrm{BH}$ of the apparent horizon    is defined in terms of the area and the quasilocal (Komar-type) angular momentum; see (\ref{PI}). For such a choice one has --- as discussed above --- the relation  $M_\mathrm{C} \approx M_\mathrm{BH}$. This is a kind of a virial relation; we shall treat its fullfilment as a test for   the selfconsistency and correctness of our numerical description.    Let us remark, that there exists    another  --- exact --- virial relation, discussed in \cite{MSH}, that can be used to check the numerical self-consistency.  

Our polytropic matter within a torus satisfies the dominant energy condition.  There is no analytic proof, but there exists numerical evidence \cite{MSH,prd2018a,prd2018b}   that the asymptotic mass $M_\mathrm{ADM}$ is not smaller than the quasilocal mass $M_\mathrm{BH}=\sqrt{\frac{S}{16\pi }+\frac{4\pi J^2_\mathrm{BH}}{S}}$.   Thus    (\ref{2}) should hold, 
\begin{equation}
M_\mathrm{ADM} \ge \sqrt{ \frac{S}{16\pi }+\frac{4\pi  J_\mathrm{BH}^2}{S}}.
 \label{3c}
\end{equation}
Obviously,  the original  Penrose inequality (\ref{1}) also holds true. 

The above statements of this subsection should be true, whenever there exist numerical solutions. The inspection of  the Table confirms this expectation.  

\subsection{On new proposals}

  Inequality (\ref{3}) uses the size measure $\tilde R(S)$ of the apparent horizon   \cite{Anglada}.  This quantity is difficult to calculate, but we can use a bound, that was shown in \cite{Anglada}, that  $\tilde R(S)$ is not larger   than      $\sqrt{10}M_\mathrm{C}$, which in turn is approximated by $\sqrt{10}M_\mathrm{BH}$.  Thus the necessary condition for the validity of (\ref{3}) reads
\begin{equation}
M_\mathrm{ADM}^2 \ge \frac{S}{16\pi }+\frac{  J_\mathrm{BH}^2}{10 M^2_\mathrm{BH}}
\label{3a}
\end{equation}
Inequality  (\ref{3a}) is valid in all our numerical  examples reported in the forthcoming Table --- compare relevant values in the column denoted as $M_\mathrm{ADM}$ with suitable entries in the last column denoted as $I_3$. This does not mean, however, that (\ref{3}) is confirmed, since (\ref{3a}) constitutes only the necessary condition.  

The inequality
 \begin{equation}
M_\mathrm{BH}^2 \ge \frac{S}{16\pi }+\frac{  J_\mathrm{BH}^2}{4 M^2_\mathrm{BH}}
\label{3b}
\end{equation}
follows directly from the definition of $M_\mathrm{BH}$, since $M_\mathrm{BH}^2 \ge \frac{S}{16\pi }$. The equality occurs for $a=J_\mathrm{BH}/m=0$.  The Table confirms that --- compare relevant values in the column denoted as $M_\mathrm{BH}$ with suitable entries in the last column denoted as $I_3$.

The quasilocal inequalities (\ref{2}) and (\ref{3}) are awkward in a sense, since they require the use of quasilocal measures of the   angular momentum.    There exists a conserved quantity related with Killing vectors, in stationary and axially symmetric quantities, that gives rise to  a distinguished (Komar-type)  quasilocal measure  of the  angular momentum. We used this fact in Section 2. Unfortunately, there is no such a quasilocal measure
  in general spacetimes.  The question arises, whether one can replace  $J_\mathrm{BH}$ by its global counterpart   $J_\mathrm{ADM}$, that is whether
\begin{equation}
M_\mathrm{ADM}^2 \ge \frac{S}{16\pi }+\frac{4\pi J^2_\mathrm{ADM}}{S}, 
\label{4}
\end{equation}
or (in a weaker formulation)
\begin{equation}
M_\mathrm{ADM}^2 \ge \frac{S}{16\pi }+\frac{ J^2_\mathrm{ADM}}{4M_\mathrm{ADM}^2},
 \label{5}
\end{equation}
 at least for stationary and asymptotically flat  spacetimes with compact material systems.
 This restriction to compact material systems is necessary, since it is easy to envisage a classical mechanical system, with an arbitrarily large angular momentum, so that both inequalities (\ref{4}) and (\ref{5}) are broken. In all examples considered below the circumferential radii of  the outermost part of tori are smaller  than 39 $M_\mathrm{ADM}$. 
 
  Let us mention here the recent work of Kopi\'nski and Tafel \cite{Tafel}, in which they  consider spacetimes  arising from a class  of perturbations of the spinless Schwarzschild geometry. These perturbations carry an angular momentum, that yields the asymptotic value $J_{\mathrm{ADM}} $.  Kopi\'nski and Tafel  prove   that (\ref{5}) is valid for such spacetimes. 
  
When using the Table, in order to test the inequality (\ref{4}),  one should compare entries of the column designated as $M_{\mathrm{ADM}} $ with relevant terms in the column denoted as $I_1$.
One can see, that (\ref{4}) is not valid for systems with heavy toroids, irrespective of the spin of the central black holes ---  see the cases H1 -- H{15}, L3MR and L5M1 -- L5M5. Let us remark, that a similar conclusion can be drawn from Table II of \cite{prd2016}, but that numerical analysis have used a perturbative approach, and therefore it is not convincing.
For lighter disks, the inequality (\ref{4}) is valid for  spins of the black hole that are not too big, but it is broken if $a$ is large enough --- see just a few examples corresponding to $a=0.9$:  L5, L10,  L15 and L20.

We have found only one counterexample to the inequality (\ref{5}) --- see the case L3MR and compare  relevant elements in the columns denoted as $M_\mathrm{ADM}$ and $I_2$.

\section{Conclusions} 
We  numerically investigate the validity of various versions of the Penrose inequality, in particular those  that include angular momentum. This is done by analyzing a stationary, axially symmetric system consisting of a black hole and a rotating polytropic torus. The original version formulated by Penrose \cite{Penrose1973} is always true.   Formulations, that bound the mass $M_\mathrm{ADM}$ by quasilocal quantities --- the area of the black hole $S$, its mass $M_\mathrm{BH}$ and angular momentum $J_\mathrm{BH}$ --- are also satisfied in our numerical solutions. We have found, however, counterexamples to   those    versions of the inequality, that are expressed in terms   of the asymptotic angular momentum $J_\mathrm{ADM}$.

\begin{table*}
\caption{\label{tab_sols}
Table. Black hole --- tori solutions. Subsequent columns contain
(from the left to the right): the solution number, the rotation law
parameter $\delta$, the black-hole spin parameter
$a$, the inner radius of the torus $r_{1}$, the outer radius of
the torus $r_{2}$, the total asymptotic mass $M_\mathrm{ADM}$,
the black hole mass $M_\mathrm{BH}$, the black hole surface $S$, the toroid angular
momentum $J_\mathrm{T}$, the total angular momentum $J_\mathrm{ADM}$ and variants of terms in various Penrose inequalities: $I_1=\sqrt{S/(16\pi)+4\pi J^2_\mathrm{ADM}/S}$, $I_2=\sqrt{S/(16\pi)+J^2_\mathrm{ADM}/(4M_\mathrm{ADM}^2)}$, $I_3=\sqrt{S/(16\pi)+J_\mathrm{BH}^2/(4M_\mathrm{BH}^2)}$.
The solutions were obtained assuming $m=1$,   $\kappa=(1-3\delta)/(1+\delta)$
and $\gamma=4/3$.}
\begin{ruledtabular}
\begin{tabular}{c c c c c c c c c c c c c}
No.      & $\delta$ & $a$      & $r_1$ & $r_2$ & $M_\mathrm{ADM}$ & $M_\mathrm{BH}$ & $S$ & $J_\mathrm{T}$ & $J_{\mathrm{ADM}} $                                & $I_1$      & $I_2$        & $I_3$\\
\hline
L1       & $-1/3$    & $-0.9$  & $6$    & $41$  & $1.100$                     & $1.001$                 & $36.19$               & $0.5204$            & $-0.3796$                     & $0.8775$ & $0.8659$ & $0.9604$\\
L2       & $-1/3$    & $-0.5$  & $6$    & $41$  & $1.100$                     & $1.003$                 & $47.26$               & $0.5008$            & $8.431\times 10^{-4}$  & $0.9697$ & $0.9697$ & $1.001$\\
L3       & $-1/3$    & $0$      & $6$    & $41$  & $1.100$                     & $1.005$                 & $50.79$               & $0.4846$            & $0.4846$                      & $1.034$   & $1.029$   & $1.005$\\
L4       & $-1/3$    & $0.5$   & $6$    & $41$  & $1.100$                     & $1.003$                 & $47.23$               & $0.4883$            & $0.9883$                      & $1.095$   & $1.068$   & $1.001$\\
L5       & $-1/3$    & $0.9$   & $6$    & $41$  & $1.100$                     & $1.000$                 & $36.17$               & $0.4947$            & $1.395$                        & $1.181$   & $1.059$.  & $0.9601$\\
\hline
L6       & $-1/7$    & $-0.9$  & $6$    & $41$  & $1.100$                     & $1.001$                 & $36.21$               & $0.4758$            & $-0.4242$                     & $0.8848$ & $0.8704$ & $0.9605$\\
L7       & $-1/7$    & $-0.5$  & $6$    & $41$  & $1.100$                     & $1.004$                 & $47.32$               & $0.4528$            & $-4.716\times 10^{-2}$ & $0.9705$ & $0.9704$ & $1.002$\\
L8       & $-1/7$    & $0$      & $6$    & $41$  & $1.100$                     & $1.006$                 & $50.87$               & $0.4326$            & $0.4326$                      & $1.029$   & $1.025$   & $1.006$\\
L9       & $-1/7$    & $0.5$   & $6$    & $41$  & $1.100$                     & $1.004$                 & $47.28$               & $0.4331$            & $0.9331$                      & $1.083$   & $1.059$   & $1.001$\\
L10     & $-1/7$    & $0.9$   & $6$    & $41$  & $1.100$                     & $1.001$                 & $36.18$               & $0.4371$            & $1.337$                        & $1.158$   & $1.0436$ & $0.9602$\\
\hline
L11     & $-1/10$  & $-0.9$  & $6$    & $41$  & $1.100$                     & $1.001$                 & $36.21$               & $0.4676$            & $-0.4324$                     & $0.8862$ & $0.8712$ & $0.9605$\\
L12     & $-1/10$  & $-0.5$  & $6$    & $41$  & $1.100$                     & $1.004$                 & $47.32$               & $0.4446$            & $-5.543\times 10^{-2}$ & $0.9707$ & $0.9706$ & $1.002$\\
L13     & $-1/10$  & $0$      & $6$    & $41$  & $1.100$                     & $1.006$                 & $50.87$               & $0.4242$            & $0.4242$                      & $1.028$   & $1.024$   & $1.006$\\
L14     & $-1/10$  & $0.5$   & $6$    & $41$  & $1.100$                     & $1.004$                 & $47.28$               & $0.4242$            & $0.9242$                      & $1.081$   & $1.057$   & $1.001$\\
L15     & $-1/10$  & $0.9$   & $6$    & $41$  & $1.100$                     & $1.001$                 & $36.18$               & $0.4277$            & $1.328$                        & $1.154$   & $1.041$   & $0.9602$\\
\hline
L16     & $0$        & $-0.9$  & $6$    & $41$  & $1.100$                     & $1.001$                 & $36.21$               & $0.4477$            & $-0.4525$                     & $0.8897$ & $0.8733$ & $0.9605$\\
L17     & $0$        & $-0.5$  & $6$    & $41$  & $1.100$                     & $1.004$                 & $47.32$               & $0.4246$            & $-7.535\times 10^{-2}$ & $0.9710$ & $0.9709$ & $1.002$\\
L18     & $0$        & $0$      & $6$    & $41$  & $1.100$                     & $1.006$                 & $50.88$               & $0.4041$            & $0.4041$                      & $1.026$   & $1.023$   & $1.006$\\
L19     & $0$        & $0.5$   & $6$    & $41$  & $1.100$                     & $1.004$                 & $47.29$               & $0.4030$            & $0.9030$                      & $1.076$   & $1.053$   & $1.001$\\
L20     & $0$        & $0.9$   & $6$    & $41$  & $1.100$                     & $1.001$                 & $36.18$               & $0.4052$            & $1.305$                        & $1.145$   & $1.035$   & $0.9602$\\
\hline
H1      & $0$        & $-0.9$  & $8$    & $20$  & $2.000$                     & $1.011$                 & $37.88$               & $5.307$              & $4.407$                        & $2.683$   & $1.403$   & $0.9756$\\
H2      & $0$        & $0$      & $8$    & $20$  & $2.000$                     & $1.079$                 & $58.49$               & $4.879$              & $4.879$                        & $2.506$   & $1.628$   & $1.079$\\
H3      & $0$        & $0.9$   & $8$    & $20$  & $2.000$                     & $1.007$                 & $37.29$               & $4.942$              & $5.842$                        & $3.499$   & $1.695$   & $0.9702$\\
\hline
H4      & $-0.2$    & $-0.9$  & $8$    & $20$  & $2.000$                     & $1.011$                 & $37.88$               & $5.362$              & $4.462$                        & $2.713$   & $1.414$   & $0.9755$\\
H5      & $-0.2$    & $0$      & $8$    & $20$  & $2.000$                     & $1.078$                 & $58.46$               & $4.938$              & $4.938$                        & $2.531$   & $1.639$   & $1.078$\\
H6      & $-0.2$    & $0.9$   & $8$    & $20$  & $2.000$                     & $1.007$                 & $37.28$               & $5.013$              & $5.913$                        & $3.539$   & $1.711$   & $0.9702$\\
\hline
H7      & $-0.4$    & $-0.9$  & $8$    & $20$  & $2.000$                     & $1.011$                 & $37.87$               & $5.406$              & $4.506$                        & $2.737$   & $1.422$   & $0.9755$\\
H8      & $-0.4$    & $0$      & $8$    & $20$  & $2.000$                     & $1.078$                 & $58.41$               & $4.988$              & $4.988$                        & $2.552$   & $1.648$   & $1.078$\\
H9      & $-0.4$    & $0.9$   & $8$    & $20$  & $2.000$                     & $1.007$                 & $37.27$               & $5.075$              & $5.975$                        & $3.574$   & $1.724$   & $0.9701$\\
\hline
H10    & $-0.6$    & $-0.9$  & $8$    & $20$  & $2.000$                     & $1.011$                 & $37.87$               & $5.436$              & $4.536$                        & $2.753$   & $1.428$   & $0.9754$\\
H11    & $-0.6$    & $0$      & $8$    & $20$  & $2.000$                     & $1.077$                 & $58.35$               & $5.032$              & $5.032$                        & $2.572$   & $1.656$   & $1.077$\\
H12    & $-0.6$    & $0.9$   & $8$    & $20$  & $2.000$                     & $1.007$                 & $37.26$               & $5.134$              & $6.034$                        & $3.608$   & $1.737$   & $0.9700$\\
\hline
H13    & $-0.8$    & $-0.9$  & $8$    & $20$  & $2.000$                     & $1.011$                 & $37.88$               & $5.437$              & $4.537$                        & $2.753$   & $1.428$   & $0.9756$\\
H14    & $-0.8$    & $0$      & $8$    & $20$  & $2.000$                     & $1.077$                 & $58.28$               & $5.074$              & $5.074$                        & $2.591$   & $1.664$   & $1.077$\\
H15    & $-0.8$    & $0.9$   & $8$    & $20$  & $2.000$                     & $1.007$                 & $37.24$               & $5.219$              & $6.119$                        & $3.657$   & $1.755$   & $0.9698$\\
\hline
L3MR & $-1/3$    & $0$      & $50$  & $75$  & $2.000$                     & $1.016$                 & $51.90$               & $9.984$              & $9.984$                        & $5.016$   & $2.695$   & $1.016$\\
L5M1 & $-1/3$    & $0.9$   & $6$    & $41$  & $1.650$                     & $1.003$                 & $36.66$               & $3.379$              & $4.279$                        & $2.647$   & $1.553$   & $0.9646$\\
L5M2 & $-1/3$    & $0.9$   & $6$    & $41$  & $2.475$                     & $1.008$                 & $37.46$               & $8.382$              & $9.282$                        & $5.445$   & $2.064$   & $0.9717$\\
L5M3 & $-1/3$    & $0.9$   & $6$    & $41$  & $3.713$                     & $1.017$                 & $38.79$               & $17.57$              & $18.47$                        & $10.55$   & $2.637$   & $0.9836$\\
L5M4 & $-1/3$    & $0.9$   & $6$    & $41$  & $5.569$                     & $1.033$                 & $41.17$               & $35.47$              & $36.37$                        & $20.11$   & $3.389$   & $1.004$\\
L5M5 & $-1/3$    & $0.9$   & $6$    & $41$  & $8.353$                     & $1.066$                 & $46.01$               & $72.76$              & $73.66$                        & $38.51$   & $4.512$   & $1.046$\\
\end{tabular}
\end{ruledtabular}
\end{table*}


\begin{thebibliography}{10}
 
\bibitem{Penrose1969} R. Penrose, Gravitational Collapse: the Role of General Relativity,   Rivista del Nuovo Cimento  {\textbf 1} 252(1969).
\bibitem{Penrose1973} R. Penrose, Naked singularities, Ann. New York Acad. Sci. \textbf{224}, 125(1973).
\bibitem{KMS} J. Karkowski,  E. Malec and Z. Swierczynski,  The Penrose inequality in spheroidal geometries, Classical and Quantum
 Gravity {\textbf 10}, 1361(1993).
\bibitem{Koc} J. Karkowski,  P. Koc and Z. Swierczynski, Penrose inequality for gravitational waves, Classical and Quantum Gravity \textbf{11},  1535(1994).
\bibitem{MNOM} E. Malec and N. O'Murchadha,  Trapped surfaces and the Penrose inequality in spherically symmetric
geometries 6931(1994), Phys. Rev. D {\textbf 49};  E. Malec, M. Iriondo and N. O'Murchadha, Constant mean curvature slices of asymptotically flat
spherical spacetimes, Phys. Rev. D {\textbf 54}, 4792(1996).
\bibitem{Hayward} S. Hayward, Inequalities Relating Area, Energy, Surface Gravity, and Charge of Black Holes, Phys. Rev. Lett. {\textbf 81}, 4557(1998).
\bibitem{Huisken} G. Huisken, T. Ilmanen, The Riemannian Penrose Inequality, Int. Math. Res. Not. \textbf{20}, 1045(1997).
\bibitem{Huisken1} G. Huisken and T. Ilmanen,  The inverse mean curvature flow and the Riemannian Penrose inequality, J. Diff. Geom., \textbf{59}, 353(2001).
\bibitem{Bray} H. Bray, Proof of the Riemannian Penrose Inequality Using the Positive Mass Theorem, J. Diff. Geom. {\textbf 59}, 177(2001).
\bibitem{Frauendiener} J. Frauendiener, On the Penrose Inequality, Phys. Rev. Lett. {\textbf 87}, 101101(2001).
\bibitem{MMS} E. Malec, M. Mars and W. Simon, On the Penrose Inequality for general horizons, Phys. Rev. Lett. {\textbf 88}, 121102(2002).
\bibitem{KM} J. Karkowski and E. Malec, The general Penrose inequality: lessons
from numerical evidence, Acta Phys. Pol. B \textbf{36}, 59(2005).
\bibitem{Mars} M. Mars, Present status of the Penrose inequality, Classical and Quantum Gravity \textbf{ 26},   193001(2009).
\bibitem{Christodoulu} D. Christodoulou, Reversible and Irreversible Transformations in Black-Hole Physics, Phys. Rev. Lett. \textbf{25}, 1596(1970).
\bibitem{Dain_Gabach}  S.\ Dain and  M.\ E.\ Gabach-Clement, Geometrical inequalities bounding angular momentum and charges in General Relativity, Living Reviews in Relativity,  \textbf{21}, 5(2018). 
\bibitem{Anglada} P. Anglada, Comments on Penrose inequality with angular momentum for outermost apparent horizons, Classical and Quantum Gravity \textbf{37}, 065023(2020).
\bibitem{Khuri} J. Jaracz and M. Khuri, Bekenstein bounds, Penrose inequalities and black hole formation, Physical Review  D \textbf{97},    124026(2018).
\bibitem{MSH} M.\ Shibata, Rotating black hole surrounded by self-gravitating torus in the puncture framework, Phys.\ Rev.\ D \textbf{76}, 064035 (2007). 
\bibitem{prd2018a} J.\ Karkowski, W.\ Kulczycki, P.\ Mach, E.\ Malec, A.\ Odrzywo{\l}ek, and M.\ Pir\'{o}g, General-relativistic rotation: Self-gravitating fluid tori in motion around black holes, Phys.\ Rev.\  D \textbf{97}, 104034 (2018).   
\bibitem{prd2018b} J.\ Karkowski, W.\ Kulczycki, P.\ Mach, E.\ Malec, A.\ Odrzywo{\l}ek, and M.\ Pir\'{o}g, Self-gravitating axially symmetric disks in general-relativistic rotation, Phys.\ Rev.\   D \textbf{97}, 104017 (2018).
\bibitem{BrandtSeidel1995} S.\ R.\ Brandt and E.\ Seidel, Evolution of distorted rotating black holes. I.\ Methods and tests, Phys.\ Rev.\ D \textbf{52}, 856 (1995). 
\bibitem{FM}   L.\ G.\ Fishbone and V.\ Moncrief, Relativistic fluid disks in orbit around Kerr black holes, Astrophys. J. {\textbf 207}, 962(1976).
\bibitem{Butterworth_Ipser} E. \ Butterworth and I.\ Ipser, Rapidly rotating fluid bodies in general relativity,  Astrophys.\ J.\ \textbf{200}, L103(1975).
\bibitem{Bardeen_1970} J.\ M.\ Bardeen, A variational principle for rotating stars in General Relativity,
Astrophys.\ J.\ \textbf{162}, 71 (1970).
\bibitem{KM2020} W. Kulczycki and E. Malec, General-relativistic rotation laws in fluid tori around spinning black holes,   Phys. Rev.  D {\textbf 101},  084016(2020). 
\bibitem{MM} P.\ Mach and E.\ Malec, General-relativistic rotation laws in rotating fluid bodies, Phys.\ Rev.\ D \textbf{91}, 124053 (2015).
\bibitem{APP2015} Jerzy Knopik, Patryk Mach and Edward Malec, General-relativistic rotation laws in rotating fluid bodies: constant linear velocity, Acta Phys.\ Pol.\ B \textbf{46}, 2451 (2015).
\bibitem{pardiso} O.\ Schenk, K.\ G\"{a}rtner, Future Generation Computer Systems 20, 475 (2004).
\bibitem{lapack} E.\ Anderson et al., LAPACK Users' Guide (SIAM, Philadelphia, 1999).
\bibitem{Tafel} J. Kopi\'nski and J. Tafel, The Penrose inequality for nonmaximal perturbations of the Schwarzschild initial data, Classical and Quantum Gravity \textbf{37}, 105006(2020).
\bibitem{prd2016} J.\ Karkowski, P.\ Mach, E.\ Malec, M.\ Pir\'{o}g, and N.\ Xie, Rotating systems, universal features in dragging and antidragging effects, and bounds of angular momentum, Phys.\ Rev.\ D \textbf{94}, 124041 (2016). 
 
\end{thebibliography}
\end{document}